\documentclass[manuscript]{aastex}

\shorttitle{Molecular Clouds Towards RCW49 and Westerlund 2}
\shortauthors{Furukawa et al.}

\begin{document}

\title{Molecular Clouds Towards RCW49 and Westerlund 2; \\
Evidence for Cluster Formation Triggered by Cloud-Cloud Collision}

\author{N. Furukawa, J. R. Dawson, A. Ohama, A. Kawamura, 
N. Mizuno, T. Onishi and Y. Fukui}
\affil{Department of Physics and Astrophysics, Nagoya University, Chikusa-ku, Nagoya, Aichi, 464-8601}

\email{naoko@a.phys.nagoya-u.ac.jp}


\begin{abstract}

We have made CO($J=2-1$) observations towards the HI\hspace{-.1em}I region RCW 49 and its ionizing source, 
the rich stellar cluster Westerlund 2, with the NANTEN2 sub-mm telescope. These observations 
have revealed that two molecular clouds in velocity ranges of $-11$ to $+9$ km s$^{-1}$ and 11 to 21 
km s$^{-1}$ respectively, show remarkably good spatial correlations with the {\it Spitzer} IRAC mid-infrared 
image of RCW 49, as well a velocity structures indicative of localized expansion around the 
bright central regions and stellar cluster. This strongly argues that the two clouds are 
physically associated with RCW 49. We obtain a new kinematic distance estimate to RCW 49 and 
Wd2 of $5.4^{ + 1.1}_{ - 1.4}$ kpc, based on the mean velocity and velocity spread of the associated gas. 
We argue that acceleration of the gas by stellar winds from Westerlund 2 is insufficient to 
explain the entire observed velocity dispersion of the molecular gas, and suggest a scenario 
in which a collision between the two clouds $\sim$4 Myrs ago may have triggered the formation of 
the stellar cluster.

\end{abstract}


\section{Introduction}

Massive stars have an enormous impact on the state of the interstellar medium 
(ISM) and on galactic evolution as a whole.
Studies of the ISM around young OB clusters provide 
valuable information on the nature of the interaction between massive stars and their 
natal environments. They may also provide vital clues to the mechanisms behind the 
clustered mode of star formation, in which nearly all massive stars are formed.

Westerlund 2 (hereafter Wd2) is an unusually rich and compact young cluster 
located close to the tangent of the Carina arm, at $l \sim 284$ degrees. With an age of 
2-3 $\times 10^{6}$ yrs \citep{pia98}, it is one of the youngest known clusters 
in the Galaxy. Estimates of the total stellar mass range 
from $\sim$1-3 $\times 10^{4} M_\odot$ \citep[][hereafter W04]{asc07,whi04}, with several 
thousand $M_\odot$ in the form of stars of 1 $\leq M \leq 120$ $M_\odot$ \citep{asc07,rau07}.
The cluster contains an extraordinary collection of hot 
and massive stars, including at least dozen O stars and two remarkable WR stars 
with strong stellar winds. These stars are ionizing a large, luminous HI\hspace{-.1em}I region, 
RCW 49 \citep{rod60}. 

The distance to Wd2 is uncertain, and values ranging from 2.8 kpc to 8.3 kpc have 
been estimated by a variety of different methods \citep[e.g.][]{asc07,rau07}. Comprehensive discussions are 
given in \citet[][hereafter C04]{chu04} and \cite{aha07}.

Recent observations with the Infrared Array Camera (IRAC) on the {\it Spitzer Space 
Telescope} have produced spectacularly detailed high-resolution mid-infrared 
images of PAH ( = polycyclic aromatic hydrocarbon) emission from RCW 49 (C04), 
as part of the GLIMPSE survey \citep{ben03}. Analysis of the IR 
point source population in the region has revealed that star formation is 
still ongoing in the nebula, with as much as $\sim$4,500 $M_\odot$ in the form of young 
stellar objects (W04). This
suggests a substantial reservoir of molecular may still remain. In addition, 
observations with the HESS telescope have revealed a TeV gamma-ray counterpart 
to Wd2, HESS J1023-575 \citep{aha07}. Possible emission scenarios 
include the interaction of cosmic ray protons accelerated in stellar wind 
shocks or supernova blast waves, with dense molecular gas. 

Molecular observations are 
crucial in pursuing the formation of 
the stellar cluster and ongoing activity in the region, and may also help 
to resolve the long-standing uncertainty in the distance to Wd2. Recently, 
\citet[][hereafter D07]{dam07} used 12CO($J=1-0$) survey data from the 
CfA 1.2 m telescope \citep{dam01} to argue for the association of 
Wd2 with a 
giant molecular cloud (GMC) on the far side of the 
Carina arm, at a kinematic distance of 6.0 kpc. This GMC 
extends almost a degree to either side of Wd2, and shows what 
appears to be a perturbed spatial and velocity structure in the vicinity of 
the cluster and HI\hspace{-.1em}I region. D07 identifies emission components at velocities 
(with respect to the local standard of rest) of $V_{{\rm LSR}} \sim16$ km s$^{-1}$ and $\sim$4 km s$^{-1}$, 
which he suggests may be parts of the $\sim$11 km s$^{-1}$ GMC that have been accelerated 
by stellar winds from the OB and WR stars. However, at 
$\sim$8.8 arcmin, the effective angular resolution is low, and the author notes 
that higher resolution observations will be needed to determine the true 
nature of the molecular gas and to investigate in detail its interaction with
the cluster. 

We here present CO($J=2-1$) observations at a resolution of 1.5 arcmin, 
taken with the NANTEN2 4 m sub-mm telescope of Nagoya University. 
Matching molecular cloud morphology to structural 
features seen by {\it Spitzer}, we vastly increase our ability to identify genuinely 
associated molecular gas, revealing valuable new information on the true 
nature of the ISM around Wd2.

\section{Observations}

Observations of the $J=2-1$ transition of $^{12}$CO were made with the NANTEN2 
4 m sub-mm telescope of Nagoya University at Atacama (4800 m above 
sea level) in Chile in February 2008. The half-power beam width of the 
telescope was 1.5 arcmin at 230 GHz. The 4 K cooled SIS mixer receiver 
provided a typical system temperature of $\sim$200 K in the single-side band at 
230 GHz, including the atmosphere toward the zenith. The spectrometer was 
an acousto-optical spectrometer with 2048 channels, providing a velocity 
coverage of 390 km s$^{-1}$ with a velocity resolution of 0.38 km s$^{-1}$ at 230 GHz. 
The data were obtained by using an OTF (=on the fly) mapping technique.  
The final pixel size and velocity channel width of the gridded data are 30 
arcsec and 0.95 km s$^{-1}$, respectively. The effective integration time per pixel 
is $\sim$2 s, and the rms noise per channel is $\sim$0.9 K.

\section{Results}

NANTEN2 CO($J=2-1$) emission over an LSR velocity range of $-100$ to 
$+100$ km s$^{-1}$ was compared with publicly available GLIMPSE images 
at 3.6, 4.5, 5.8 and 8.0 microns. We find tight morphological 
correlations between the IR nebula and molecular gas components 
peaking at velocities of $\sim$16 km s$^{-1}$ and $\sim$4 km s$^{-1}$, strongly suggesting 
that these clouds are directly associated with the HI\hspace{-.1em}I region and stellar 
cluster. In addition, we find similarly strong evidence of association 
with a hitherto un-noted velocity components with peak velocities of 
between $-9$ and 0 km s$^{-1}$. 

\begin{figure}
\epsscale{.80}
\plotone{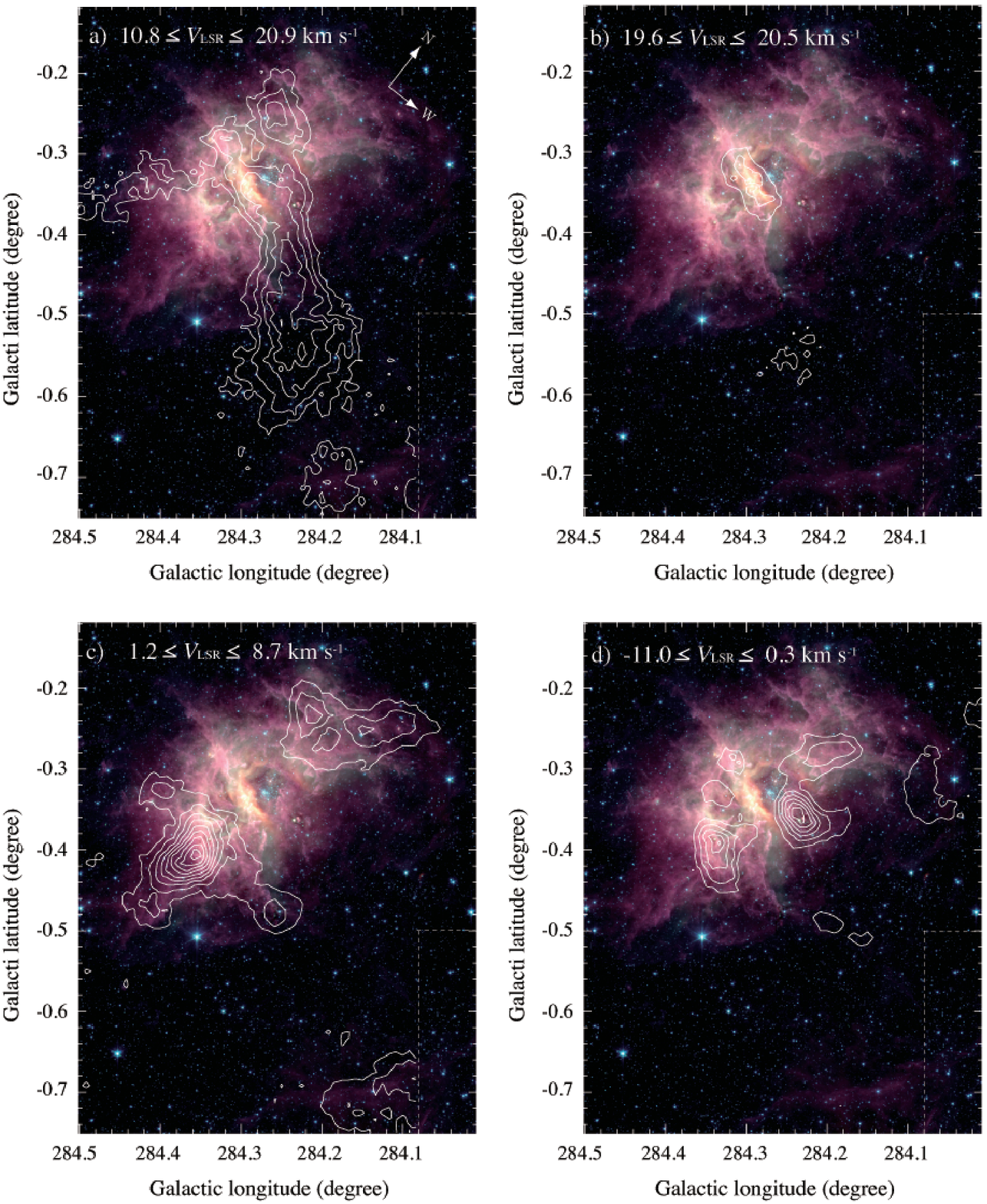}
\caption{Spitzer/IRAC GLIMPSE 3 colour image of RCW 49 overlaid with 
$^{12}$CO($J=2-1$) contours. The 3.6, 4.5 and 8.0 micron wavebands are shown 
in blue, green and red, respectively. Panels (a), (b), (c) and (d) show CO 
($J=2-1$) emission integrated over velocity ranges of $10.8 \leq V_{{\rm LSR}} \leq 20.9$ km s$^{-1}$,
$19.6 \leq V_{{\rm LSR}} \leq 20.5$ km s$^{-1}$, 
$1.2 \leq V_{{\rm LSR}} \leq 8.7$ km s$^{-1}$ and 
$-11.0 \leq V_{{\rm LSR}} \leq 0.3$ km s$^{-1}$, 
with contour levels of 20 + 10 K km s$^{-1}$, 
5 + 2 K km s$^{-1}$, 17 + 15 K km s$^{-1}$ and 30 + 15 K km s$^{-1}$ respectively. \label{fig1}}
\end{figure}

Figure \ref{fig1} shows the GLIMPSE image overlaid with CO integrated intensity 
contours. Figure \ref{fig1}a shows the CO distribution between 11 and 21 km s$^{-1}$, 
corresponding to the $\sim$16 km s$^{-1}$ cloud identified by D07. The cloud is 
elongated from the Northeast to the Southwest, extending over the range 
($284.15 \leq l \leq 284.34$) and ($-0.63 \leq b \leq -0.20$). Emission removed from the 
main body of the cloud to the Southeast, at ($l$, $b$) $\sim$ (284.40, $-0.33$), and 
Southwest, at ($l$, $b$) $\sim$ (284.19, $-0.71$), represents molecular gas identified 
by D07 as part of the $\sim$11 km s$^{-1}$ GMC. The Southeastern edge of the 16 km s$^{-1}$ 
cloud between (284.33, $-0.28$) and (284.27, $-0.40$) shows a strong 
correlation with the bright filamentary ridge of mid-infrared emission 
$\sim$2 arcmin Southeast of Wd2. This ridge is especially bright in the 4.5 
micron band, which is dominated by the hydrogen recombination line Br, 
indicating the presence of highly ionized gas (C04). The CO emission 
shows signs of localized velocity perturbation coincident with this 
ridge, with the peak velocity offset to $\sim$19 km s$^{-1}$ (illustrated in figure \ref{fig1}b). 
This strongly suggests a physical association between the molecular gas 
and RCW 49. In addition, the 16 km s$^{-1}$ cloud as a whole shows some sign of 
depleted emission towards Wd2 itself, consistent with the presence of 
the evacuated region within 1' $\sim$ 2' of the cluster.  

Figure \ref{fig1}c shows the CO distribution between 1 and 9 km s$^{-1}$, corresponding 
to the $\sim$4 km s$^{-1}$ cloud identified in D07. Emission at this velocity 
range is entirely absent within a radius of $\sim$3' of Wd2. 
At larger distances from the cluster, molecular gas fans out to the North 
and South, showing an excellent large-scale correlation with extended 
shape of the IR nebula. We also note the close correspondence between 
the edge of the CO emission and the curved rim of filamentary IR emission 
at around (284.22, $-0.29$). 

Figure \ref{fig1}d shows the CO distribution between $-11$ and 0 km s$^{-1}$. This emission 
contains components with peak velocities between $-9$ and 0 km s$^{-1}$. The 
brightest clump in this velocity range is centered on ($l$, $b$) $\sim$ (284.23, $-0.36$), 
 $\sim$3' Southwest of Wd2, at a peak velocity of $-4$ km s$^{-1}$, and 
is coincident with a region of ongoing massive star formation in the nebula
(W04). To the Northwest, CO emission neatly skirts the edge of the evacuated region 
around the stellar cluster.
Emission in this velocity range is newly identified here as associated 
with RCW 49 and Wd2, however, it is not distinctly separated from the 4 
km s$^{-1}$ cloud. 
We hereafter refer to all emission in the range $-11$ to $+9$ km s$^{-1}$ as 
the 0 km s$^{-1}$ cloud, based on the intensity-weighted mean velocity of the entire 
complex.

The velocities of the associated components allow us to obtain kinematic 
distances. 
Using the rotation curve of \cite{bra93}, 
we obtain of distances of 6.5, 5.2 and 4.0 kpc, for 
velocities of 16, 4 and $-4$ km s$^{-1}$ respectively. The intensity-weighted mean 
velocity of all associated emission is 6 km s$^{-1}$, which corresponds to a 
distance of 5.4 kpc. In light of the large velocity spread and associated 
uncertainty, we 
adopt a conservative distance estimate of $5.4 
^{+ 1.1}_{- 1.4}$ kpc. In the interests of consistency this value is used for 
all emission components throughout this paper.

The mass of the associated molecular gas is estimated from $^{12}$CO($J=1-0$) 
data from the NANTEN Galactic Plane Survey \citep{miz04}. Although 
the resolution is lower that the present dataset ($\sim$4'), the relevant 
components are readily identified. We adopt a conversion factor ('X factor') 
of $2.0 \times 10^{20}$ cm$^{-2}$ K km s$^{-1}$. 
This results 
in molecular hydrogen masses, $M(\mathrm{H}_2)$, of $9.1 \pm 4.1 \times 10^4 M_\odot$ and $8.1 \pm 3.7 \times 
10^4 M_\odot$ for the 16 and 0 km s$^{-1}$ clouds, where the uncertainties propagate 
through from the distance estimate. The higher resolution CO ($J=2-1$) data 
were also used to derive a virial mass for the 16 km s$^{-1}$ cloud 
\citep[see e.g.][]{kaw98}, which was found to be almost identical 
to the luminosity-based value. We did not attempt to derive virial 
masses for the 0 km s$^{-1}$ cloud, since its velocity distribution is complex 
and apparently un-relaxed.

D07 argues that the 16 km s$^{-1}$ cloud 
lies behind RCW 49 and the 4 km s$^{-1}$ component in front, based on an 
analysis of HI absorption spectra seen against the bright continuum 
background. We concur with his reasoning and note that the absorption 
spectrum also shows an additional peak at around  $\sim$-4 km s$^{-1}$, 
corresponding to the brightest concentration of gas in our most 
blue-shifted velocity range, placing it too in front of the nebula 
(see figure 2 of D07 and related discussion). This result is confirmed 
on inspection of optical data from the ESO Digitized Sky Survey. 
Good agreement is seen between optical obscuration and the 0 km s$^{-1}$ cloud, 
whereas the 16 km s$^{-1}$ cloud shows no obvious correlation with dark features.

\section{Discussion}

In the above we have demonstrated the association of Wd2/RCW 49 with a 
significant mass of molecular gas spanning 
almost 30 km s$^{-1}$, consisting of a foreground component at $\sim$0 km s$^{-1}$ and a background component at $\sim$16 km s$^{-1}$. This velocity dispersion is an order of 
magnitude too large for the clouds to be gravitationally bound. Yet 
their physical association with RCW 49 implies that 
their spatial separation must be no more than  $\sim$40 pc (the linear dimension of the nebula at 5.4 kpc). 

An obvious interpretation is that the molecular gas is 
expanding away from the central cluster, presumably driven by the energy 
output of the massive stars within it. This is 
favored 
by D07, and is consistent with the placement of the various velocity 
components along the line of sight. However, while it is true that the 
molecular gas towards the IR nebula does show some signs of localized 
perturbation due to the cluster's influence, we are cautious about 
interpreting the global gas configuration according to this scenario. 

\begin{figure}
\plotone{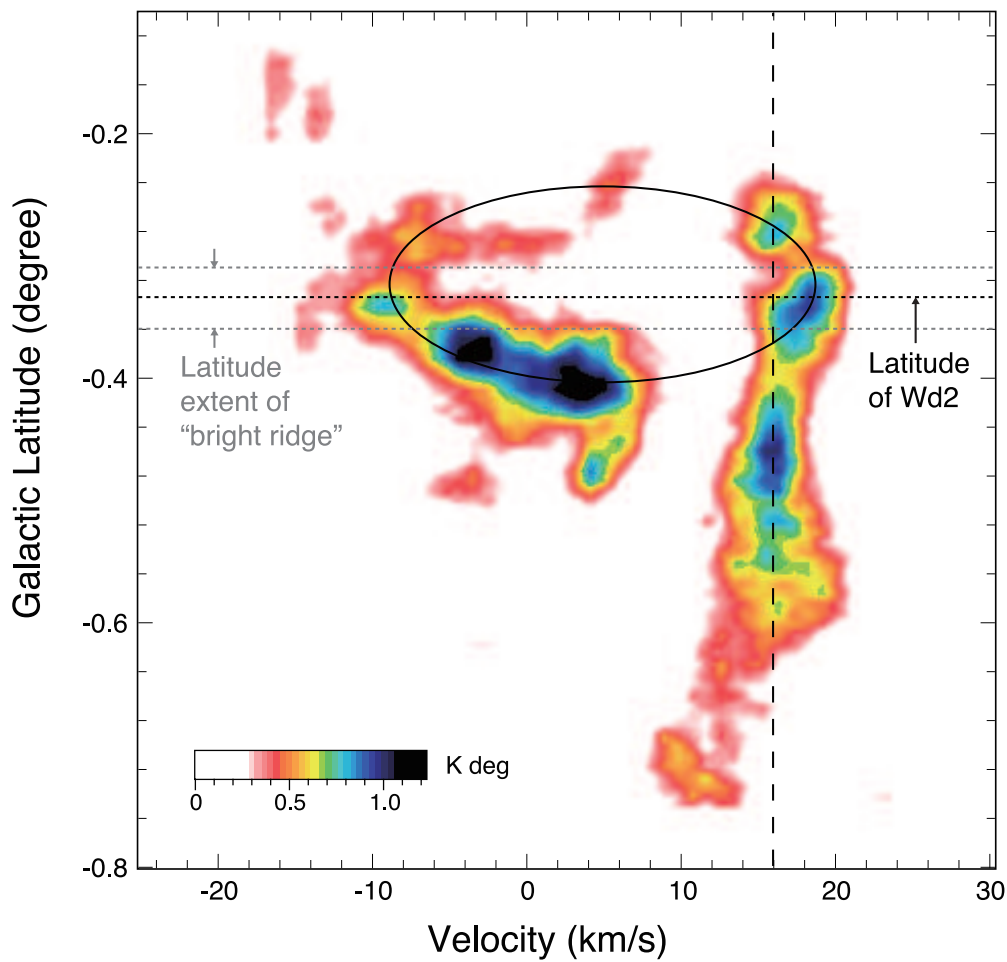}
\caption{Velocity vs Galactic latitude diagram for $^{12}$CO($J=2-1$) emission 
integrated over a longitude range of 284.2 to 284.4 degrees. Emission 
between $-11$ and $+8$ km s$^{-1}$ and between $+12$ and $+21$ km s$^{-1}$ respectively 
corresponds to the 0 and 16 km s$^{-1}$ clouds referred to in the text. 
Solid black circle: velocity signature consistent 
with expansion around the central cluster and/or bright central regions 
of RCW 49. Dashed line: illustrates the constant velocity of the 
16 km s$^{-1}$ cloud outside the region of direct influence of the cluster. 
Dotted lines mark the latitudes of Wd2 and the bright ridge of 
IR emission referred to in the text.\label{fig2}}
\end{figure}

Figure \ref{fig2} shows a CO($J=2-1$) velocity-latitude diagram. 
The 16 and 0 km s$^{-1}$ clouds are 
well separated in $b$-$v$ space. 
Gas located 
directly in the line of sight of the IR nebula shows a velocity signature 
that is consistent with expansion around the central cluster. This is 
especially true of the 0 km s$^{-1}$ cloud, which shows the fastest approaching 
material located in the direction of the bright central regions, surrounded 
by gas with less extreme line-of-sight velocities. 
The top of the 16 km s$^{-1}$ cloud also shows evidence for interaction, 
localized to the immediate vicinity of the bright IR ridge, 
and far less pronounced. However, the majority of the cloud remains at a constant 
velocity of 16 km s$^{-1}$, and 
extends as far as 12' ( $\sim$20 pc) outside 
the outer boundary of RCW 49; 
presumably well outside the 
range of influence of the cluster. It is therefore difficult to explain 
the entire 16 km s$^{-1}$ cloud as a velocity-perturbed clump of the extended 11 
km s$^{-1}$ GMC, as originally suggested by D07. 

Furthermore, we may show that any scenario that attempts to explain the 
entire cloud velocity spread by cluster-driven expansion alone is 
energetically problematic. 
Considering the line-of-sight 
velocity difference of the 0 and 16 km s$^{-1}$ clouds relative to a systemic velocity of 
6 km s$^{-1}$, and using the cloud masses estimated above, we obtain an 
estimated K.E. of $\sim$1.2 $\pm 0.5 \times 10^{50}$ erg. 
Significantly changing the underlying 
assumptions in this simplistic calculation (e.g. varying the systemic velocity, limiting the included mass 
to gas within the boundary of RCW 49, considering a 3D expansion velocity 
based on the measured $V_{{\rm LSR}}$ extremes, etc.) does not greatly affect 
this value, which remains  $\sim$10$^{50}$ erg. The $v^2$ dependence of K.E. combined 
with the large velocity spread of the associated material ensures this. 

The total mechanical luminosity available 
from stellar winds in Wd2 is estimated by \cite{rau07} as $3.6 \times 10^{51}$ erg. (The 
authors note that no supernova should yet have occurred in the cluster's lifetime). 
The required kinetic energy is therefore at least several \% of the total 
available energy from the cluster. Although in ideal adiabatic wind 
bubbles 20\% of the wind luminosity is transferred 
to the expanding neutral shell \citep{wea77}, more realistic 
numerical models suggest that no more than a few \% of the initial wind 
energy ends up as neutral gas KE \citep{art07}. In addition, 
the small solid angle ($<4\pi$) subtended by the clouds 
reduces the energy available to them. We therefore conclude 
that while it is not impossible that stellar wind-driven expansion  
is responsible for the entire velocity separation of the molecular clouds, 
the energy requirements are uncomfortably tight.  

The implication 
is that a certain proportion of 
the cloud velocities are systemic: i.e. unrelated to the dynamical 
interaction of the complex with Wd2 and RCW 49. 
We of course only have information on a single velocity component, 
and the fact that the 0 and 16 km s$^{-1}$ clouds are receding from each other along 
the line of sight does not necessarily imply that they were once 
spatially coincident. Nevertheless, rather than postulating a non-interactive 
close approach between the two clouds, 
we may consider whether 
direct interaction between them might have played a role in the 
formation of a cluster as rich as Wd2. Collisions between molecular clouds 
can lead to gravitational instability in the dense, shocked gas, resulting 
in 
triggered star formation \citep[see][and reference within]{elm98}. 
Such collisions between molecular clouds are 
presumably rare. Very roughly, the average time between collisions is  $\sim$$1/\sigma nv$, where $n$ 
is the cloud number density, $\sigma$ is the cross sectional area and $v$ is 
the cloud-cloud r.m.s. velocity dispersion. Assuming that the Milky Way 
contains  $\sim$4000 giant molecular clouds with $\sigma\approx30^2\pi$ pc$^2$, $v\approx5$ km s$^{-1}$, distributed in a 100 parsec-thick ring between $3 < R < 8$ kpc, 
this value is 
$\sim$6 $\times$ 10$^8$ yr. However, this interval 
will be considerably smaller in the spiral arms where the cloud density 
is higher, and non-random motions may increase the cloud collision rate. 
For a relative velocity of $\sim$10 km s$^{-1}$ the time taken to cover a distance of 
40 pc (the assumed separation of the clouds) is $\sim$4 Myr, which is highly 
consistent with the estimated age of Wd2 of 2-3 Myr. We therefore conclude 
that triggered formation of the Wd2 cluster via cloud-cloud collision is a 
viable - and intriguing - scenario. We also note that the star formation 
efficiency, defined as the ratio of the cluster mass to the molecular mass 
of the 16 and 0 km s$^{-1}$ clouds, is  $\sim$5\%, which is consistent with typical Galactic values.

We finally return briefly to the question of whether Wd2/RCW 49 and the 
molecular clouds presented here may also be associated with the larger 11 
km s$^{-1}$ GMC, as suggested by D07. 
The GMC extends almost a degree to the East and West of Wd2/RCW 49, and with the 
exception of a small sub-component at (284.42, $-0.34$), does not fall directly 
along the line of sight of the nebula. As described above, the 16, and 0 km s$^{-1}$ 
clouds of this study are not well explained as clumps accelerated from an 
original systemic velocity of $\sim$11 km s$^{-1}$. The required energy input is uncomfortably 
large, and the large offset between much of the 16 km s$^{-1}$ cloud and RCW 49 argues 
against acceleration of the entire cloud by the cluster. The new identification 
of associated gas with negative velocities also drags the intensity 
weighted mean velocity 
from 11 km s$^{-1}$ to 6 km s$^{-1}$. Finally 
we also note that the mass of 
the directly associated components 
seems to be sufficient to form such a rich cluster, even without invoking the 
larger reservoir of gas contained in the 11 km s$^{-1}$ GMC. Nevertheless, we do not 
completely rule out a loose association with the molecular clouds presented 
here. We also note that the revised distance estimate of 5.4$^{+ 1.1}_{- 1.4}$ kpc 
presented in this study is consistent with D07's value of 6.0 kpc.

Finally, we note that presence of the TeV gamma ray source HESS J1023-575 suggests 
that the present clouds are good candidates for gamma ray production via 
interaction with cosmic ray protons, and deserve further detailed scrutiny.

\section{Summary}

\begin{enumerate}
\item We have identified molecular gas associated with Wd2 and RCW 49 based on 
strong correlations between CO($J=2-1$) emission and mid-infrared data from the 
{\it Spitzer Space Telescope} GLIMPSE survey. The molecular gas spans an unusually wide 
velocity range of  $\sim$30 km s$^{-1}$, with two distinct complexes with mean velocities of 
16 and 0 km s$^{-1}$. 
\item The intensity weighted mean velocity of the entire mass of associated gas 
is $\sim$6 km s$^{-1}$, and this is used to obtain a new kinematic distance estimate to 
the cluster and HI\hspace{-.1em}I region, of 5.4$^{+ 1.1}_{- 1.4}$ kpc. This value is unavoidably 
crude due to the large velocity dispersion of the system. The total molecular 
hydrogen mass of the associated gas is estimated as $1.7 \pm 0.8 \times 10^5 M_\odot$. 
\item The molecular gas velocity structure shows evidence for expansion around 
Wd2 and the central parts of RCW 49. However, consideration of the energetics, 
and also the extent of the 16 km s$^{-1}$ cloud well outside of the region of 
influence of RCW 49, argues against the observed velocity dispersion arising 
entirely as a result of this interaction.
\item We suggest that a collision between the two molecular clouds may have 
triggered the formation of Wd2 $\sim$4 Myr ago.
\item The clouds presented in this study are viable as a target for TeV gamma 
ray production via interaction with cosmic ray protons.
\end{enumerate}

\acknowledgments

We are grateful to Tom Dame for his feedback and comments, which have
led to a significant improvement of this manuscript. NANTEN2 is an
international collaboration between 10 universities, Nagoya, Osaka
Prefecture, Cologne, Bonn, Seoul National, Chile, New South Wales, MacQuarie, Sydney
and Zurich. This work is financially supported in part by a Grant-in-Aid
for Scientific Research (KAKENHI) from the MEXT and
from JSPS, in part, through the core-to-core program (No. 17004). This work is based in part
on archival data obtained with the {\it Spitzer Space Telescope}, which
is operated by the Jet Propulsion Laboratory, California Institute of Technology under
a contract with NASA.

\clearpage

\clearpage

\end{document}